\documentclass[twocolumn,floatfix,showpacs,showkeys,pra,aps,a4paper]{revtex4}

 \usepackage{bm}
 \usepackage{amsmath}
 \usepackage{amssymb}
 \usepackage{latexsym}
 \usepackage{amsfonts}
 \usepackage{epsfig}
 \usepackage{color}
 \definecolor{darkblue}{rgb}{0,0,.5}
 \definecolor{dgreen}{rgb}{0.0, 0.5, 0.0}
 \definecolor{orange}{rgb}{1.0, 0.5, 0.0}
 \usepackage[linktocpage, colorlinks=true ,linkcolor=darkblue, citecolor=darkblue ]{hyperref}
 \usepackage[all]{hypcap}

 \newcommand{\ket}[1]{\left|#1\right>}
 \newcommand{\bra}[1]{\left<#1\right|}
 \newcommand{\expval}[1]{\left< #1 \right>}
 
 \newcommand{\braket}[2]
 {\left<#1|#2\right>}
 \newcommand{\nn}{\nonumber\\}
 
 \newcommand{\f}[1]{\mbox{\boldmath$#1$}}

 \newcommand{\bea}{\begin{eqnarray}}
 \newcommand{\ea}{\end{eqnarray}}
 \newcommand{\eea}{\end{eqnarray}}
 \newcommand{\ord}{{\cal O}}
 \newcommand{\trace}[1]{{\rm Tr}\left\{ #1 \right\}}
 \newcommand{\traceB}[1]{{\rm Tr_B}\left\{ #1 \right\}}
 \newcommand{\traceS}[1]{{\rm Tr_S}\left\{ #1 \right\}}
 \newcommand{\abs}[1]{{\left| #1 \right|}}
 
 \newcommand{\HS}{H_{\rm S}}
 \newcommand{\HB}{H_{\rm B}}
 \newcommand{\HI}{H_{\rm SB}}
 \newcommand{\RS}{\rho_{\rm S}}
 \newcommand{\RB}{\rho_{\rm B}}

 \newcommand{\sgn}[1]{{\rm sgn}\left( #1 \right)}

 \newcommand{\bin}[2]{\left(\begin{array}{c}#1\\#2\end{array}\right)}

\begin{document}

\title{The speed of Markovian relaxation towards the ground state}

\author{Malte Vogl}
\author{Gernot Schaller}\email{schaller@itp.physik.tu-berlin.de}
\author{Tobias Brandes}

\affiliation{Institut f\"ur Theoretische Physik, Technische Universit\"at Berlin, Hardenbergstr. 36, 10623 Berlin, Germany}

\begin{abstract}
For sufficiently low reservoir temperatures, it is known that open quantum systems
subject to decoherent interactions with the reservoir
relax towards their ground state in the weak coupling limit.
Within the framework of quantum master equations, this is formalized by the Born-Markov-secular (BMS) approximation, where
one obtains the 
system 
Gibbs state with the reservoir temperature as a stationary state.
When the solution to some problem is encoded in the (isolated) ground state of a system Hamiltonian, decoherence can therefore 
be exploited for computation.
The computational complexity is then given by the scaling of the relaxation time with the system size $n$.

We study the relaxation behavior for local and non-local Hamiltonians that are coupled dissipatively 
with local and non-local operators to a bosonic bath in thermal equilibrium.
We find that relaxation is generally more efficient when coherences of the density matrix 
in the system energy eigenbasis are taken into account.
In addition, the relaxation speed strongly depends on the matrix elements of the coupling operators between
initial state and ground state.

We show that Dicke superradiance is a special case of our relaxation models and can thus 
be understood as a coherence-assisted relaxation speedup.
\end{abstract}

\pacs{
03.65.Yz, 
31.15.xp, 
03.67.-a 
}

\keywords{Lindblad form, stationary state, dissipation}

\maketitle


\section{Introduction}

Since the seminal papers by Shor~\cite{shor1997} and Grover~\cite{grover1997} the interest in quantum computation is unbroken 
and has given rise to plenty of research~\cite{nielsenchuang2000}.

However, the superiority of quantum computation relies on the fragile quantum coherence of states in the system 
of interest, which is sensible to interactions with the environment~\cite{unruh1995}. 
The process of loosing coherence, the so called decoherence, has yet to be fully understood~\cite{giulinizeh1996,schlosshauer2008}.
Various attempts are constantly made to describe the process of decoherence for special systems analytically~\cite{breuer2004a}. 
The hope is that one can overcome 
the devastating influence of decoherence
by error correction codes~\cite{karasik2007} -- 
or by developing new models of computation
such as measurement-based~\cite{raussendorf2001a,raussendorf2003a}, holonomic~\cite{pachos2001a}, ground-state~\cite{mizel2001,mizel2002} or adiabatic~\cite{farhi2001} 
quantum computation.

In this paper we describe the (unitary and non-unitary) effects of decoherence by means of a Markovian master equation, 
namely the Born-Markov-secular (BMS) approach~\cite{breuerpetruccione2002}. 
The resulting master equation is an effective evolution equation for the reduced density matrix of the system.
Such master equations are valid in the weak-coupling and long-time limit provided the reservoir relaxation time is 
significantly faster than the system dynamics.
The BMS master equation is always of Lindblad type \cite{lindblad1976a} and thus preserves positivity of the
reduced system density matrix at all times.
Beyond this it can be shown, that for a thermalized reservoir with reservoir temperature $\beta$ the equilibrated system Gibbs state 
\bea\label{Egibbs}
\rho_{\rm th}=\frac{e^{-\beta \HS}}{\traceS{e^{- \beta \HS}}}\,,
\eea
with the same temperature $\beta$ is a stationary state of the BMS master equation~\cite{breuerpetruccione2002}.
When we label the energies of the system Hamiltonian with a unique ground state by $E_0 < E_1 \le \ldots \le E_{N-1}$ for
an $N$-dimensional system Hilbert space, the probability to find the stationary state in the systems ground state
is given by
\bea\label{Ebeta}
P_0=\frac{e^{-\beta E_0}}{\sum\limits_{i=0}^{N-1}e^{-\beta E_i}} \ge \frac{1}{1+N e^{-\beta (E_1-E_0)}}\,,
\eea
where the lower bound is a rather crude estimate.
Assuming a system composed of $n$ qubits (such that $N=2^n$), this implies a large probability to 
reach the ground state by dissipative relaxation when the temperature $k_{\rm B} T = \beta^{-1}$ scales as
\bea\label{Etempscaling}
k_{\rm B} T(n) \propto \frac{(E_1-E_0)}{n}\,.
\eea

This remarkably mild dependence of the required temperature on the system size $n$ is not spoilt by the
fundamental energy gap $E_1-E_0$:
In fact, it is possible to 
encode the solution to interesting problems in the ground state of local Hamiltonians
with a fundamental energy gap that does not even scale with the system size $n$ at all.
Since such Hamiltonians can easily be constructed for the NP-complete problems Exact Cover 3 \cite{farhi2001,schallerschuetzhold2006} 
and 3-SAT \cite{znidaric2005a}, it does not come as a surprise that also the solution to the 
prominent factoring problem can be encoded in the ground state of a two-body local Hamiltonian \cite{schallerschuetzhold2010}.

The robustness of the ground state against decoherence at low temperatures is already exploited in various computation schemes
(see e.g.~\cite{farhi2001,childsetal2001,mizel2009,mizel2001,mizel2002}).
However, here we are aiming at solving problems using nothing but decoherence-induced relaxation \cite{cirac2009}.
Given a corresponding system Hamiltonian, the question for the speed of this relaxation process arises naturally.
The dependence of the relaxation speed on the system size defines the computational 
complexity of this cooling algorithm.

%
%

In this paper we investigate the scaling behaviour of the relaxation speed for several time-independent (and scalable) Hamiltonians.
The paper is organized as follows:
In section \ref{SMethod} we introduce our methods, which we apply to nonlocal and local Hamiltonians
in sections \ref{sec:relaxI} and \ref{sec:sigmax}, respectively.


\section{Method}\label{SMethod}

We separate the Hamiltonians for the various problems in this paper as $H = \HS +\HI + \HB$,
where $\HS$ and $\HB$ only act on the system and bath Hilbert space, respectively.
In contrast, the interaction Hamiltonian can be decomposed as
\bea
 \HI = \lambda \sum_{\cal A} A_{\cal A} \otimes B_{\cal A}\,,
\eea
with the small dimensionless coupling parameter $\lambda \ll 1$ and the system ($A_{\cal A}$) and bath ($B_{\cal A}$) coupling operators.
By a suitable redefinition~\cite{breuerpetruccione2002} we can always assume hermitian coupling operators
$A_{\cal A} = A_{\cal A}^{\dagger}$ and $B_{\cal A} = B_{\cal A}^{\dagger}$.
We define the bath-correlation functions as
\bea
    C_{\cal AB}(\tau) \equiv \traceB{e^{i \HB \tau}B_{\cal A} e^{-i \HB \tau} B_{\cal B} \RB^0} = C_{\cal AB}^*(-\tau)\,,
\eea
and introduce their even and odd Fourier transforms
\bea
  \gamma_{\cal AB}(\omega)&\equiv& \int\limits_{-\infty}^{+\infty}C_{\cal AB}(\tau)e^{i\omega\tau}d\tau\,,\nn
  \sigma_{\cal AB}(\omega)&\equiv& \int\limits_{-\infty}^{+\infty}C_{\cal AB}(\tau)\sgn{\tau}e^{i \omega \tau}\nn
   &=&-\frac{i}{\pi} {\cal P}\int\limits_{-\infty}^{+\infty}\frac{\gamma_{\cal AB}(\bar\omega)}{\bar\omega-\omega}d\bar\omega\,,
\eea
where ${\cal P}$ denotes the Cauchy principal value.
For the transformation into the interaction picture, it is useful to expand the system coupling operators into
eigenoperators of the system Hamiltonian
\bea\label{Eeigop}	
A_{\cal A} &=& \sum_\omega A_{\cal A}(\omega) = \sum_\omega A_{\cal A}^\dagger(\omega) = A_{\cal A}^\dagger
\eea
which obey by definition
$\left[ \HS, A_{\cal A}(\omega)\right] = -\omega A_{\cal A}(\omega)$, 
where $\omega$ denotes all energy differences of the system Hamiltonian.
Such a decomposition can for example be achieved in the system energy eigenbasis $H_s \ket{a} = E_a \ket{a}$ by 
using $A_{\cal A}(\omega) = \sum_{ab} \delta_{(E_b - E_a), \omega}\ket{a} \bra{a} A_{\cal A} \ket{b} \bra{b}$.
Performing the Born-Markov and the secular approximation, where the latter amounts to the neglect of the resulting
oscillating terms~\cite{breuerpetruccione2002}, we arrive at a Lindblad
form master equation for the system density matrix
\bea\label{Elindblad_rwa}
\dot{\RS} &=& -i \left[\HS, \RS(t)\right]\nn
&&-i \left[\frac{\lambda^2}{2i}\sum_{\omega} \sum_{\cal AB} \sigma_{\cal AB}(\omega) A_{\cal A}^\dagger(\omega) A_{\cal B}(\omega), \RS(t)\right]\nn
&&+\lambda^2\sum_{\omega}\sum_{\cal AB} \gamma_{\cal AB}(\omega)\times\nn
&&\times\left[
A_{\cal B}(\omega) \RS A_{\cal A}^\dagger (\omega)
- \frac{1}{2}  \left\{A_{\cal A}^\dagger(\omega) A_{\cal B}(\omega), \RS(t)\right\}\right]\,,\nn
\eea
where $\sigma_{\cal AB}(\omega)$ is anti-hermitian (leading to a hermitian Lamb-shift Hamiltonian) 
and $\gamma_{\cal AB}(\omega)$ is positive semidefinite for each $\omega$.
The Lindblad form of this equation is a sufficient condition~\cite{lindblad1976a,breuerpetruccione2002} for preserving positivity of the density
matrix for all times and coupling constants $\lambda$.
When we phrase these results in the system energy eigenbasis, the quantum master equation takes the
form (with $\rho_{ab} \equiv \bra{a}\rho\ket{b}$)~\cite{schaller2008a}
\bea\label{Equantum}
\dot\rho_{ab} &&= -i (E_a-E_b) \rho_{ab} -i \sum_c \left[\bar\sigma_{ac} \rho_{cb} - \bar\sigma_{cb} \rho_{ac}\right]\nn
&&+ \sum_{cd} \left[\bar\gamma_{ac,bd} \rho_{cd}-\frac{1}{2} \bar\gamma_{cd,ca} \rho_{db} - \frac{1}{2} \bar\gamma_{cb,cd} \rho_{ad}\right]\,,
\eea
where the damping coefficients equate to
\bea
\bar\gamma_{ab,cd} &=& \lambda^2\sum_{\cal AB}\gamma_{\cal AB}(E_b-E_a)\delta_{E_d-E_c, E_b-E_a}\nn 
&&\times\bra{a}A_{\cal B}\ket{b}\bra{c}A_{\cal A}\ket{d}^*
\eea
and the Lamb-shift terms are explicitly given as
\bea
\bar\sigma_{ab} &=& \frac{\lambda^2}{2i} \sum_c \sum_{\cal AB} \sigma_{\cal AB}(E_a-E_c) \delta_{E_a,E_b}\nn 
&&\times \bra{c} A_{\cal A} \ket{a}^* \bra{c} A_{\cal B} \ket{b}\,.
\eea
We will refer to Eq.~(\ref{Equantum}) as the quantum master equation in the following.

\subsubsection*{Dephasing model}

As a simple example for the quantum master equation, let us consider a pure dephasing model, where the coupling operator
commutes with the system Hamiltonian.
This implies that $\HS$ and $A$ can be diagonalized simultaneously $A\ket{a}=d_a \ket{a}$, 
and the coefficients acquire an especially simple form
\bea
\bar\gamma_{ab,cd}=\lambda^2 \gamma(0) \delta_{ab}\delta_{cd} d_a d_c\,,\qquad
\bar\sigma_{ab}=\frac{\lambda^2}{2i} \sigma(0) \delta_{ab} d_a^2\,,
\eea
such that we obtain for the matrix elements of the density matrix
\bea
\dot\rho_{ab}&=&\biggl\{-i\left[\left(E_a-E_b\right) + \frac{\lambda^2}{2i} \sigma(0) \left(d_a^2-d_b^2\right)\right]\nn
&&-\frac{\lambda^2\gamma(0)}{2} \left(d_a-d_b\right)^2\biggr\} \rho_{ab}\,,
\eea
which leads to an exponential decay of off-diagonal matrix elements that correspond 
to different eigenvalues of the coupling operator $A$ (recall that $\sigma(0)$ is imaginary)
\bea
\rho_{ab}(t) \propto \exp\left\{-\frac{\lambda^2\gamma(0)}{2} \left(d_a-d_b\right)^2 t\right\}\,.
\eea

\subsubsection*{Rate Equation}

It is evident from Eq.~(\ref{Equantum}) that for a non-degenerate spectrum of the system Hamiltonian the BMS dynamics of the diagonals (populations)
\bea\label{Eclassic}
\dot\rho_{aa} &=& +\sum_b \bar\gamma_{ab,ab} \rho_{bb} - \left(\sum_c \bar\gamma_{ca,ca}\right) \rho_{aa}
\eea
completely decouples from the dynamics of the off-diagonals (coherences) and is also unaffected by the
Lamb-shift terms.
Therefore -- compatible initial conditions provided -- 
the coherences may be completely neglected from the considerations.

In reality, it can be expected that exact degeneracies of ideal model Hamiltonians may be broken by perturbations.
In addition, as shown before, dephasing noise may also lead to the decay of coherences, such that the above equation can be well motivated.
We will refer to Eq.~(\ref{Eclassic}) as rate equation in the following.
Note however, that even for a non-degenerate spectrum with smallest level splitting $\Delta \epsilon$,
the secular approximation is only applicable for times $t > \Delta \epsilon^{-1}$, which should be kept
as an additional overhead in mind when discussing the scaling of relaxation. For a non-degenerate two-level system, rate equation and quantum master equation coincide.

\subsubsection*{General Assumptions}

Throughout the paper, we will assume a bosonic bath $\HB=\sum_k \omega_k \left(b_k^\dagger b_k + \frac{1}{2}\right)$ constantly held
at thermal equilibrium.
For simplicity, we will consider single-operator coupling $\HI = \lambda A\otimes \sum_k \left[ h_k b_k + h_k^* b_k^\dagger\right]$,
such that we may omit the indices and the Fourier transform of the bath correlation function is given by~\cite{schaller2008a}
\bea
\gamma_{\cal AB}(\omega) \equiv \gamma(\omega) &=& \frac{g(\abs{\omega})}{\abs{1-e^{-\beta \omega}}}\,,
\eea
where we have assumed a thermalized bath with inverse temperature $\beta$.
It should be noted that for large systems, one will in general need many different coupling operators to 
avoid non-ergodic behavior (many stationary states)~\cite{breuerpetruccione2002}.
Regarding the system coupling operators, we note that as
$\rho(t)=e^{{\cal L} t}\rho^0$ and 
${\cal L} \propto A_{\cal A}^2$, a scaling of the 
coupling operators has a direct effect on the time-dependence of the density matrix.
In order to perform a fair comparison between models with different coupling operators, we therefore concentrate -- unless noted otherwise -- on operators 
with a strictly bounded spectrum (i.e. $\|A_{\cal A}\| \leq 1$, where the spectral norm for a hermitian operator
is defined as $\|A_{\cal A}\| \equiv \sqrt{\lambda_{\rm max}^2}$).
In the following, we will consider models with a unique ground state denoted by $\ket{w}$ throughout the paper.
Furthermore, to be able to compare numerical with analytical solutions, we define the relaxation 
time $t_{\rm relax}$ for a given initial state as the time needed, 
to achieve a ground state population of $P_{\rm ground}=\bra{w}\bar\rho\ket{w}=0.9$ in the stationary state $\bar \rho$.


\section{Nonlocal Coupling Operators}\label{sec:relaxI}

All coupling operators in this section will -- an implementation based on $n$ qubits assumed --
be nonlocal in the sense that they act on all qubits simultaneously.
In addition, most of the following problems will have the symmetry that the dynamics is completely
contained in the subspace spanned by $\ket{w}$ and $\ket{s}$, where 
\bea
\ket{s} = \frac{1}{\sqrt{N}} \bigotimes\limits_{i=1}^n (\ket{0}+\ket{1})_i = \frac{1}{\sqrt{N}} \sum_{i=0}^N \ket{i}
\eea
is the superposition of all computational basis states, and $\ket{w}$ is the distinguished ground state.
An orthonormal basis for this subspace is given by
\bea\label{Ebasis2}
\ket{v_1}&=&\ket{w}\,,\nn
\ket{v_2}&=&\frac{\ket{s}-\braket{w}{s}\ket{w}}{\sqrt{\left\|\ket{s}-\braket{w}{s}\ket{w}\right\|}} 
= \frac{1}{\sqrt{N-1}} \sum_{a\neq w} \ket{a}\,.
\eea
We will consider a system Hamiltonian with the unique ground state $\ket{w}$ and exponentially many degenerate
excited states, which in basis~(\ref{Ebasis2}) can be written as
\bea\label{Ehamoracle}
\HS = \Delta E\left[\f{1} - \ket{w}\bra{w}\right] = 
\left(\begin{array}{cc}
0 & 0\\
0 & \Delta E
\end{array}
\right)
\,,
\eea
where $\ket{w}$ may for example be a marked item in a database~\cite{roland_cerf2002} or encode
the solution to a problem, such that the Hamiltonian may act as an oracle~\cite{grover1997}.
A classical minimization algorithm would find the solution $\ket{w}$ in the worst case after $N$ trials of this oracle.
Accordingly, we have $E_0=0$ and $E_1 = \ldots = E_{N-1}=\Delta E > 0$.
In order to get a large overlap between the Gibbs state~(\ref{Egibbs}) and the ground state, we 
require low temperatures 
\bea\label{Etemp1}
e^{\beta \Delta E} = \ord(N)\,,
\eea
where $N=2^n$ denotes the dimension of the Hilbert space, such that the required temperature
must scale inversely with the system size, compare Eqn.~(\ref{Etempscaling}).


\subsection{A single Projector}\label{sec:singlepro}

We consider a single coupling operator with an interaction Hamiltonian of the form 
\bea\label{Ecoupss}
A &=& \ket{s} \bra{s} 
= \bigotimes\limits_{\ell=1}^n \frac{\f{1}_\ell+\sigma^x_\ell}{2}\nn 
&=& \frac{1}{N}\left(\begin{array}{cc}
1 & \sqrt{N-1}\\
\sqrt{N-1} & N-1
\end{array}
\right)\,.
\eea
As it is a projector, the operator $A$ has eigenvalues zero and one.
%
%
We introduce the rate equation variables
\bea\label{Evarclass}
z_1^{\rm re} &\equiv& \rho_{ww}\,,\qquad z_2^{\rm re}\equiv \sum_{a\neq w} \rho_{aa}
\eea
and the quantum master equation variables variables
\bea\label{Evarquant}
z_1^{\rm qe} &\equiv& \rho_{ww}\,,\qquad z_2^{\rm qe}\equiv \sum_{a,b\neq w} \rho_{ab}\,,
\eea
with which we can both write Eqs.~(\ref{Eclassic}) and~(\ref{Equantum}) in the form
\bea\label{Ematrix}
\left(\begin{array}{c}
\dot z_1\\
\dot z_2
\end{array}\right) =
M
\left(\begin{array}{c}
z_1\\
z_2
\end{array}\right)\,,
\eea
where $M$ is a two by two matrix. 
Note that trace conservation implies that $\dot z_1^{\rm re}+\dot z_2^{\rm re}=0$, but
$z_1^{\rm qe}+z_2^{\rm qe}$ is not conserved. 
The formal solution of Eq.~(\ref{Ematrix}) is given by
\bea\label{Esolution}
\f{z}(t)=e^{M t} \f{z^0}\,,
\eea
where $\f{z^0}$ denotes the initial conditions. 
As the solution is unknown, only one initial condition that does not favor the ground state can be prepared in the rate equation approach. Here 
\bea\label{Eclassinit}
z_1^0 &=& \frac{1}{N}\,,\qquad
z_2^0 = \frac{N-1}{N}\,,
\eea
which corresponds to a diagonal density matrix with equal diagonal elements -- or simply the Gibbs state~(\ref{Egibbs}) at infinite temperatures.
In the quantum master equation the additional initial state
\bea\label{Equantinit}
z_1^0 &=& \frac{1}{N}\,,\qquad
z_2^0 = \frac{(N-1)^2}{N}
\eea
is also preparable.
This state corresponds to a $\ket{s}\bra{s}$ initial density matrix, as can be seen from Eq.~(\ref{Evarquant}).


\subsubsection{Rate Equations}\label{seq:sym_classic}

For the rate equations we obtain with variables~(\ref{Evarclass}) from Eq.~(\ref{Eclassic})
the matrix
\bea\label{eq:sym_class}
M^{\rm re} &=& \frac{\lambda^2}{N^2}\begin{pmatrix}
                                    -\gamma(-\Delta E)(N-1) & \gamma(\Delta E)\\
				    \gamma(-\Delta E)(N-1) & -\gamma(\Delta E)
                                 \end{pmatrix}\,,
\eea
Without knowledge of the solution $\ket{w}$, we can prepare the initial condition~(\ref{Eclassinit}). 
Solving Eq.~(\ref{Esolution}) with the matrix~(\ref{eq:sym_class}) we find
\bea 
z_1(t)=\frac{1}{(N+e^{\beta \Delta E}-1)}\biggl\{ e^{\beta \Delta E} - \frac{1}{N}\bigl[(N-1)\nn\times(e^{\beta \Delta E}-1)e^{-t\frac{\lambda^2}{N^2} g(\Delta E) \frac{(N+e^{\beta \Delta E}-1)}{(e^{\beta \Delta E}-1)}}\bigr]\biggr\}\,,
\eea
such that the steady state~(\ref{Egibbs}) will be reached after time $t \gg \tau_{\rm re}$, where
\bea\label{eq:taure}
\tau_{\rm re} \equiv \frac{N^2 \left(e^{\beta \Delta E} - 1\right)}{\lambda^2 g(\Delta E) \left(N + e^{\beta \Delta E} - 1\right)}
\stackrel{e^{\beta \Delta E}\to N}{\Longrightarrow} \ord\left(N^2\right)\,,
\eea
as can be seen from the exponent.
This is surprisingly inefficient.
In addition, it must be kept in mind that when we assume the exact 
degeneracies of Eq.~(\ref{Ecoupss}) to be lifted by some perturbation,
the rate equations become valid after times larger than the largest inverse level splitting, 
which would lead to an additional computational overhead.
This overhead will be significant, as the splitting must be exponentially small: The $2^n-1$ non-degenerate excited states
must fit into the range $\left[\Delta E-\delta, \Delta E+\delta\right]$ with $\delta \ll \Delta E$
when the Hamiltonian~(\ref{Ehamoracle}) should be approximately valid.
However, alternatively we may assume that dephasing noise leads to the decay of coherences and thereby motivate
the simplified rate equations after a significantly shorter timescale.


\subsubsection{Quantum Master Equations}\label{seq:sym_quantum}

With the coherences, relaxation is more efficient. 
With variables~(\ref{Evarquant}) we obtain from Eq.~(\ref{Equantum})
the evolution matrix
\bea\label{eq:sym_quant}
M^{\rm qe} &=& \lambda^2 \frac{(N-1)^2}{N^2}\begin{pmatrix}
                                    \frac{-\gamma(-\Delta E)}{(N-1)} & \frac{+\gamma(+\Delta E)}{(N-1)^2}\\
				    \frac{+\gamma(-\Delta E)}{(1)} & \frac{-\gamma(+\Delta E)}{(N-1)}
                                 \end{pmatrix}\,,
\eea
see also Appendix~\ref{Acoherenceone}.
For both initial conditions~(\ref{Eclassinit}) and~(\ref{Equantinit}), we obtain a decay into the steady state for times
\bea\label{eq:taume}
t \gg \tau_{\rm me} \equiv \frac{N^2 \tanh\left(\frac{\beta \Delta E}{2}\right)}{\lambda^2 g(\Delta E) (N-1)} 
\stackrel{e^{\beta \Delta E}\to N}{\Longrightarrow} \ord\left(N\right)\,,
\eea
which is not faster than classical minimization. 
However, for the previous initial condition~(\ref{Eclassinit}),
the overlap between steady state and ground state becomes exponentially small.
This can be understood since the first choice~(\ref{Eclassinit}) is not fully contained in the subspace spanned by the
dyadic product of~(\ref{Ebasis2}), such that only an exponentially small part of this initial condition will relax
towards the Gibbs state.
In contrast, the initial condition~(\ref{Equantinit}) corresponds to $\ket{s}\bra{s}$ and is thus 
fully contained within the relevant subspace.
In this case, relaxation approaches the Gibbs state~(\ref{Egibbs}).
This problem of the evolution of the quantum master equation~(\ref{Equantum}) towards a non-thermal state can be 
cured by using coupling operators that mediate transitions out of the two-dimensional subspace~\cite{breuerpetruccione2002}.
Note that the evolution matrices for both methods~(\ref{eq:sym_class}) and~(\ref{eq:sym_quant}) 
coincide for $N=2$, that is for one qubit $n=1$. 


\subsection{An indirect channel}\label{sec:mixedpro}

As a slight modification of the previous coupling operator, we consider
\bea
A = \eta \left(\ket{s}\bra{s}+\ket{w}\bra{w}\right)\,.
\eea
In order to have $\|A\| \leq 1$, we may choose $\eta=\frac{N}{N+1}$, which converges to one for large $N$.


With variables~(\ref{Evarclass}) Eq.~(\ref{Eclassic}) leads to the evolution matrix $M=\frac{N^2}{(N+1)^2} M^{\rm re}$, where $M^{\rm re}$ is given in (\ref{eq:sym_class}).
Therefore, with initial conditions~(\ref{Eclassinit}), the rate equation relaxation time will scale as $t \sim \ord(N^2)$, which is the same as in Eq.~(\ref{eq:sym_class}).


With variables~(\ref{Evarquant}) the evolution equation corresponding to the quantum master equation~(\ref{Equantum})
can be written as $M=\frac{N^2}{(N+1)^2} M^{\rm me}$, with $M^{\rm me}$ as given in Eq.~(\ref{eq:sym_quant}).
Again, we have the same scaling of the relaxation time as in Eq.~(\ref{eq:sym_quant}) - $\ord(N)$.
The dependences of the relaxation time on the number of states for both cases are depicted in Fig.~\ref{kopplungen_vergleich}.


\subsection{A direct channel}\label{sec:directpro}

Here we will demonstrate that the runtime can become independent on the system size when the coupling operator mediates
a direct transition between the initial state and the ground state.
In this section we focus on this by considering the coupling operator
\bea
A = \eta \left(\ket{s}\bra{w}+\ket{w}\bra{s}\right)\,,
\eea
where it is obvious that $\bra{s} A \ket{w} = \ord\{1\}$, since in order to have $\|A\| \leq 1$, 
one may choose the prefactor $\eta = \frac{\sqrt{N}}{1+\sqrt{N}}$.


Using the variables~(\ref{Evarclass}) we can write Eq.~(\ref{Eclassic}) with the
evolution matrix $M=\frac{N}{(\sqrt{N}+1)^2}M^{\rm re}$.
Therefore, the scaling of the rate equation dissipation time is given by $t \sim \ord(N)$.


With the variables~(\ref{Evarquant}) we obtain for Eq.~(\ref{Equantum}) the evolution matrix $M=\frac{N}{(\sqrt{N}+1)^2}M^{\rm me}$.
For initialization with an $\ket{s}\bra{s}$ density matrix as in Eq.~(\ref{Equantinit}) one obtains 
a large overlap of the ground state and the stationary state, and
the relaxation time scales as $t \sim \ord{(1)}$!
These scaling behaviors are summarized in Fig.~\ref{kopplungen_vergleich}.


\subsection{Hadamard Coupling}\label{sec:relaxII}

Inspired by the nonlocal representation of the previous coupling operator~(\ref{Ecoupss})
we choose a product of Hadamard gates as coupling operator
\bea
A = \bigotimes_{\ell=1}^n \frac{\sigma^z_\ell+\sigma^x_\ell}{\sqrt{2}} = \bigotimes_{\ell=1}^n {\cal H}_\ell\,.
\eea
The eigenvalues of the operator $A$ are given by $-1$ and $+1$ (with exponentially large degeneracies) 
and its matrix elements in the computational basis read
\bea
\bra{a} A \ket{b} &=& \prod_{\ell=1}^n \bra{a_{\ell}} {\cal H}_{\ell} \ket{b_{\ell}} = \prod_{\ell =1}^n  \frac{(-1)^{a_{\ell} b_{\ell}}}{\sqrt{2}}\nn
&=& \frac{(-1)^{\sum\limits_{\ell=1}^n a_\ell b_\ell}}{\sqrt{N}} \equiv \frac{(-1)^{a \circ b}}{\sqrt{N}}\,,
\eea
where $ a \circ b \equiv \sum_{\alpha=1}^n a_\alpha b_\alpha$ denotes the Hamming weight (the number of ones in the binary decomposition) of the bitwise product of $a$ and $b$.


\subsubsection{Rate Equations}\label{classical_eq}

\begin{figure}[ht]
\includegraphics[clip=true, width=0.45\textwidth]{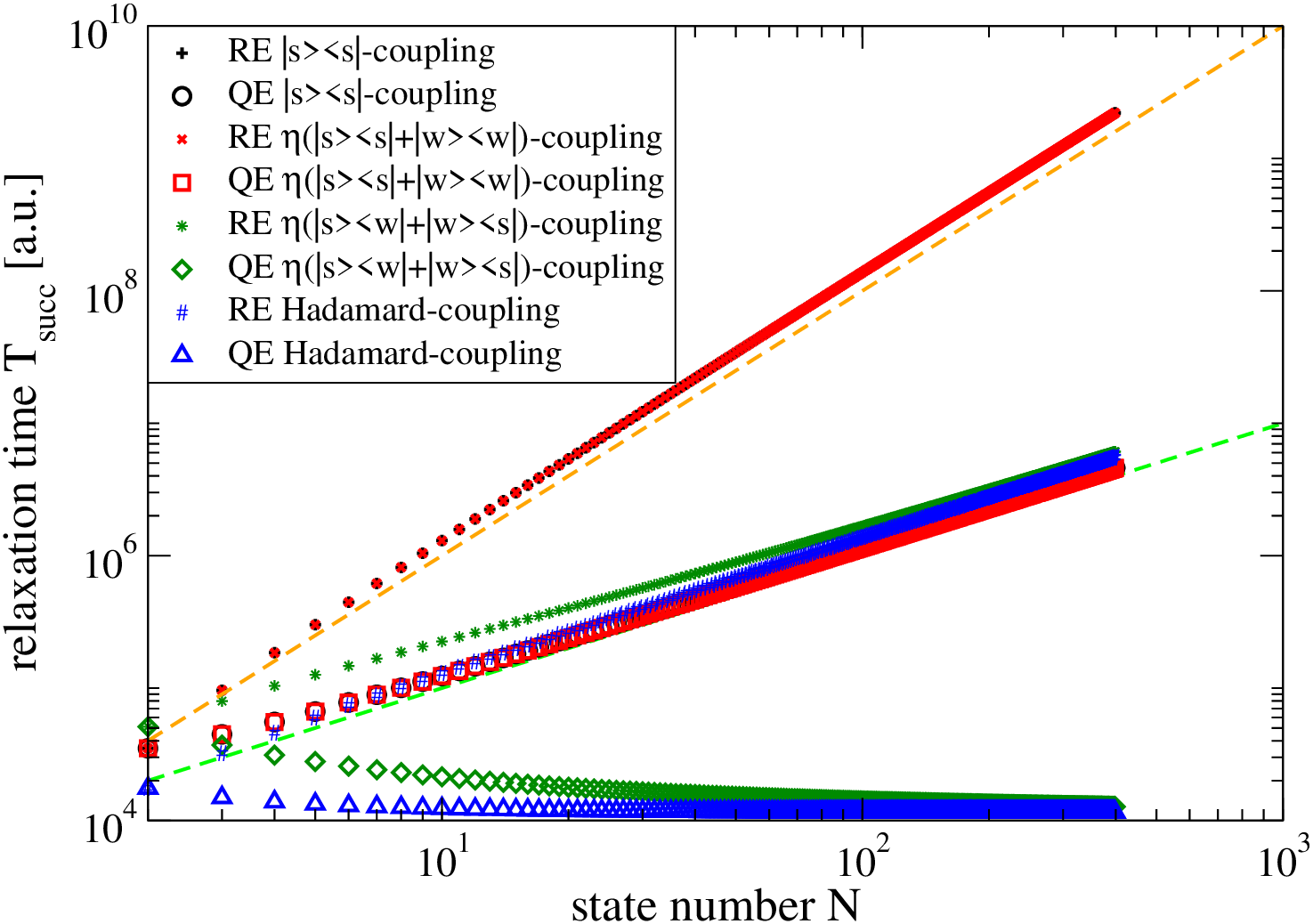}
\caption{\label{kopplungen_vergleich}
[Color Online]
Scaling of the runtime to achieve $P_{\rm ground}=0.9$ versus the number of states $N$.
For each $N$, the inverse temperature $\beta$ has been adapted such that the ground state probability
of the Gibbs state~(\ref{Egibbs}) was $P_{\rm ground}=0.95$.
For all couplings one observes a clear speedup for the master equation results (hollow symbols) in comparison
to the rate equation results (symbols).
The rate equation solution for the Hadamard coupling (blue $\#$-symbols) is of the same order as the quantum master equation solution
of Subsec.~\ref{seq:sym_quantum}. 
The runtime for the quantum master case the Hadamard coupling (blue triangles up) is only finite if $\ket{0 \cdots 0}$ 
is chosen as the solution.
By construction, the quantum master equation runtime is always finite for the operator of Sec.~\ref{sec:directpro} (green diamonds). 
Dashed lines denote the asymptotic scalings $T\propto N$ and $T\propto N^2$, respectively.
Other parameters have been chosen as $g(\abs{\Delta E})= 2$, $\Delta E=1$, and $\lambda =0.01$.
}
\end{figure}

In analogy to Subsec.~\ref{sec:relaxI}, we obtain due to the different matrix elements $\abs{\bra{a} A \ket{b}}^2=1/N$
with the variables~(\ref{Evarclass}) the evolution matrix $M = N M^{\rm re}$, cf. Eq.~(\ref{eq:sym_class}).
For the initial equal distribution~(\ref{Eclassinit}) we obtain relaxation into the steady state~(\ref{Egibbs}) after
\bea
t \gg \frac{\tau_{\rm re}}{N}
\stackrel{e^{\beta \Delta E}\to N}{\Longrightarrow} \ord\left(N\right)\,,
\eea
with $\tau_{\rm re}$ given in Eq.~(\ref{eq:taure}), which corresponds to the classical complexity.
Note that the steady state is thus reached in similar time as for the quantum case of Subsec.~\ref{sec:singlepro}, 
as can be seen from Fig.~\ref{kopplungen_vergleich}. 


\subsubsection{Quantum Master Equations}\label{quantum_eq}

Due to the different matrix elements of our coupling, we cannot close Eq.~(\ref{Equantum}) with 
the variables~(\ref{Evarquant}).
Instead, one achieves closure with the variables
\bea\label{Evarhadamard}
z_1 = \rho_{ww}\,,\qquad
z_2 = \sum_{ab \neq w} (-1)^{w \circ a + w \circ b} \rho_{ab}.
\eea
Insertion into Eq.~(\ref{Equantum}) yields an evolution equation of type $\f{\dot{z}}=M \f{z}$ with $M = N M^{\rm me}$, cf. Eq.~(\ref{eq:sym_quant}) and
see Appendix~\ref{ArelaxII} for details.
Again, we do now have different possibilities of initialization without knowing the actual ground state $w$. 

For example, by using a diagonal density matrix as initial condition as in Eq.~(\ref{Eclassinit}) we obtain relaxation after times
\bea
t \gg \frac{\tau_{\rm me}}{N}
\stackrel{e^{\beta \Delta E}\to N}{\Longrightarrow} \ord\left(1\right)\,,
\eea
with $\tau_{\rm me}$ as given in Eq.~(\ref{eq:taume}), but the stationary state always has exponentially small overlap with the ground state.

Even when we initialize the density matrix with $\ket{s}\bra{s}$ (see Appendix~\ref{ap:overlap} for the details), 
the overlap of the stationary state with the ground state $\ket{w}$ will be exponentially small -- 
unless the ground state is given by $\ket{w}=\ket{0}\otimes\ldots\otimes\ket{0}$.
In the latter case, relaxation also becomes extremely fast, since the relaxation time becomes independent
of the system size.
This is due to the fact that the Hadamard coupling provides a direct channel from the initial state 
towards the ground state, as
$\ket{s}={\cal H}^{\otimes n} \ket{0\ldots 0}$.

\section{Local coupling operators}\label{sec:sigmax}


\subsection{Climbing a Hamming ladder}

Here we consider a Hamiltonian that gives an energy penalty $E_\alpha$ to all states that have Hamming distance
$\alpha$ to the solution $\ket{\omega}$
\bea
\HS = \sum_{\alpha=0}^n E_\alpha \sum_{a \in {\cal H}_\alpha} \ket{a}\bra{a}\,,\qquad 
A = \eta \sum_{\ell=1}^n \sigma^x_\ell\,,
\eea
where ${\cal H}_\alpha$ denotes the subspace of all states with Hamming distance $\alpha$ to the solution $\ket{w}$.
The Hamming distance is a relative distance measure and is defined as the number of bitflips needed to transform one computational basis state to another.
Obviously, it ranges between 0 (equal states) and $n$ (inverted states).
The above coupling operator mediates transitions between states with Hamming distance one.
In order to constrain the eigenvalues of $A$ between $-1$ and $+1$, we need to choose $\eta=\frac{1}{n}$.
For the ease of manipulation, we order all states in the computational basis with respect
to their Hamming distance to the solution, such that we can assume that 
\mbox{$E_0 \le E_1 \le \ldots \le E_{n-1} \le E_n$}. 
There are $\bin{n}{\alpha}$ states with Hamming distance $\alpha$ to the solution $\ket{w}$,
which also gives the degeneracy of $E_\alpha$.
In the following, we will use that for each state in ${\cal H}_\alpha$, there are 
$(n-\alpha)$ single-bitflips that lead to ${\cal H}_{\alpha+1}$ and $(\alpha)$ single-bitflips that lead to ${\cal H}_{\alpha-1}$. 

As an example of this, consider $n=10$ spins and the ground state $\ket{0000000000}$.
Then, the spin configuration $\ket{1011010101} \in {\cal H}_6$. 
There are six bitflips leading to ${\cal H}_5$, namely all bitflips on ones.
In addition, the four possible bitflips on zeros lead to ${\cal H}_7$.
This generalizes to arbitrary solution states, as the Hamming distance is a relative measure.

Note that already the simplest classical algorithm for finding the ground state is quite effective, 
as one only has to flip single bits of an $n$-bit bitstring and calculate the corresponding energy:
If the energy decreases, one keeps the modified bitstring configuration, whereas in the opposite case one
tries to flip another bit.
Thus, one finds the ground state classically after at most $n$ bit flips.

Assuming an equidistant level spacing $E_\alpha = \alpha \Delta E$, we see that the condition on the temperature
to yield a large overlap between the ground state and the Gibbs state is significantly weaker
\bea\label{Etemp2}
e^{\beta \Delta E}=\ord\{n\}
\eea
than in Eq.~(\ref{Etemp1}).


\subsubsection{Rate Equations}\label{sec:sigmax_classical}

We introduce the variables 
\bea\label{Evarhamrate}
z_{\alpha}^{\rm re} \equiv \sum_{j \in {\cal H}_{\alpha}} \rho_{jj}\,.
\eea
Then, we obtain from Eq.~(\ref{Eclassic})
\bea\label{Eclasshamming}
\dot z_0^{\rm re} &=& - \lambda^2\eta^2 (n) \gamma(E_0-E_1) z_0^{\rm re}\nn
&& + \lambda^2\eta^2 (1) \gamma(E_1-E_0) z_1^{\rm re}\,,\nn
&\vdots&\nn
\dot z_{\alpha}^{\rm re} &=& + \lambda^2\eta^2 (n-\alpha+1) \gamma(E_{\alpha-1}-E_{\alpha}) z_{\alpha-1}^{\rm re}\nn
        && - \lambda^2\eta^2 \big[(\alpha) \gamma(E_{\alpha}-E_{\alpha - 1})\nn
	&& + (n-\alpha) \gamma(E_{\alpha}-E_{\alpha + 1})\big] z_{\alpha}^{\rm re}\nn 
        &&+ \lambda^2\eta^2 (\alpha+1) \gamma(E_{\alpha + 1}- E_{\alpha}) z_{\alpha+1}^{\rm re}\,,\nn
&\vdots&\nn
\dot z_n^{\rm re} &=& +\lambda^2\eta^2 (1) \gamma(E_{n-1}-E_n) z_{n-1}^{\rm re} \nn
&&- \lambda^2\eta^2 \left[(n) \gamma(E_n-E_{n-1})\right] z_n^{\rm re}\,.
\eea
When we consider equidistant spacings $E_\alpha-E_{\alpha-1} \equiv\Delta E$, the eigenvalues of the associated matrix are given by
\bea
\lambda_\alpha &=& - \alpha \lambda^2\eta^2 g(\Delta E) \coth\left(\frac{\beta \Delta E}{2}\right)\,,\nn
\eea
where $\alpha \in \{0,1,2,\ldots,n\}$ such that we obtain an efficient (polynomial) scaling of the relaxation time  $t \propto \eta^{-2} = n^2$ in this case.


\subsubsection{Quantum Master Equations}\label{sec:sigmax_quantum}

\begin{figure}[ht]
\includegraphics[clip=true, width=0.45\textwidth]{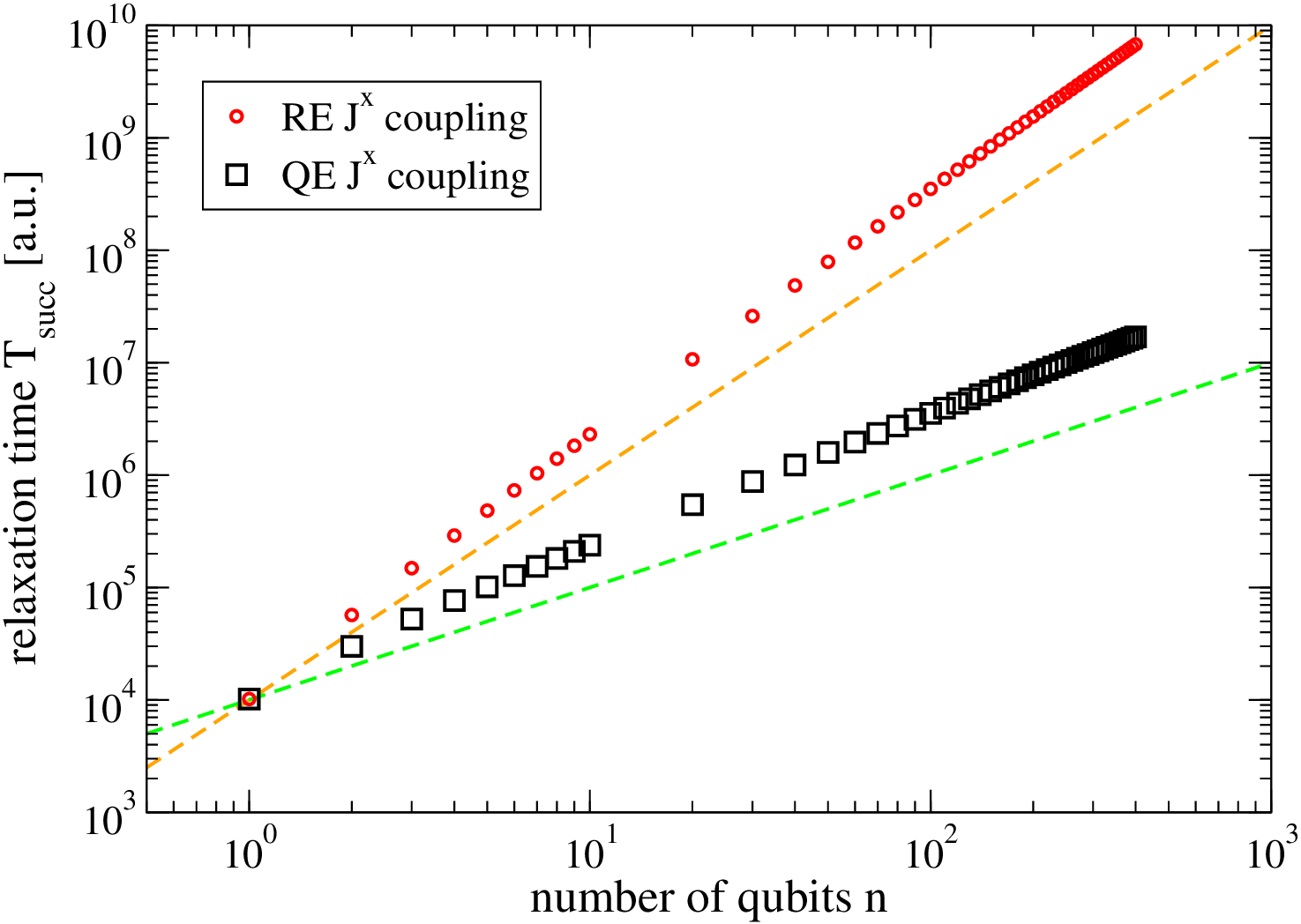}
\caption{\label{Fbitflip_vergleich}
[Color Online]
Scaling of the runtime to achieve $P_{\rm success}=0.9$ versus the number of qubits $n$.
For each $n$, the inverse temperature $\beta$ has been adapted such that the ground state
occupancy was $P_{\rm ground}=0.95$.
One observes a clear speedup for the quantum case (black squares) in comparison to 
the rate equation case (red circles), where only the populations are considered.
Dashed lines denote the asymptotic scalings $T\propto n$ and $T\propto n^2$, respectively.
Other parameters have been chosen as
$g(\abs{\Delta E})= 2$, $\Delta E=1$, and $\lambda =0.01$ .
}
\end{figure}

We introduce the variables 
\bea\label{Evarhamquant}
z_\alpha^{\rm qe} \equiv \sum_{ab \in {\cal H}_\alpha} \rho_{ab}\,.
\eea
Then, we obtain from Eq.~(\ref{Equantum})
\bea\label{Equanthamming}
\dot z_0^{\rm qe} &=& - \lambda^2\eta^2 n \gamma(E_0-E_1) z_0^{\rm qe} + \lambda^2\eta^2 \gamma(E_1-E_0) z_1^{\rm qe}\,,\nn
&\vdots&\nn
\dot z_\alpha^{\rm qe} &=& + \lambda^2\eta^2 (n-\alpha+1)^2 \gamma(E_{\alpha-1}-E_\alpha) z_{\alpha-1}^{\rm qe}\nn
&&- \lambda^2\eta^2 \biggl[\gamma(E_\alpha-E_{\alpha-1}) \alpha (n-\alpha+1)\nn 
&&+ \gamma(E_\alpha-E_{\alpha+1}) (n-\alpha) (\alpha+1)\biggr] z_\alpha^{\rm qe}\nn
&&+ \lambda^2\eta^2 (\alpha+1)^2 \gamma(E_{\alpha+1} - E_\alpha) z_{\alpha+1}^{\rm qe}\,,\nn
&\vdots&\nn
\dot z_n^{\rm qe} &=& +\lambda^2\eta^2 \gamma(E_{n-1}-E_n) z_{n-1}^{\rm qe}\nn
 &&- \lambda^2\eta^2 \gamma(E_n-E_{n-1}) z_n^{\rm qe}\,.
\eea
To make a statement about the efficiency of this algorithm, we use numerical solutions of the coupled system of differential equations. 
However, due to the tridiagonal structure of the system, standard solvers are quite efficient. 
Therefore, we can easily compute the success time, the time necessary to reach a population of 0.9 in the ground state, for up to 400 qubits. 
The comparison as depicted in Fig.~\ref{Fbitflip_vergleich} shows the clear speed-up from $\ord(n^2)$ to $\ord(n)$ for the quantum case. 


\subsection{Example: Dicke Superradiance}

The question of relaxation speed has also been raised in the field of quantum optics, since it is
connected with the energy flux radiated by the system into the reservoir.
We have observed that with coherences the relaxation speed is quadratically enhanced.
In the field of quantum optics, this is already known as Dicke superradiance, which exemplifies
the broader validity of our approach.
There, coupled relaxation equations of tridiagonal type also arise naturally:

The Hamiltonian of Sec.~\ref{sec:sigmax} is a generalization of the multi-mode Dicke Hamiltonian 
\bea\label{Ehamdicke}
H_{\textmd{Dicke}}&=& \frac{\omega_0}{2} J^z + \lambda J^x \otimes \sum_k (h_k b_k^{\dagger}+ h_k^* b_k)\nn
&&+ \sum_k \omega_k b_k^{\dagger} b_k\,,
\eea
where we have $J^\alpha \equiv \sum_{\ell=1}^n \sigma^\alpha_\ell$.
The phenomenon of superradiance as first pointed out by Dicke~\cite{dicke1957}
describes the process of coherent radiance by a system of $n$ two-level systems, that
are initially collectively excited.
The radiance, contrary to classical emission, has an intensity $I \sim n^2$ and  a 
delay time $t_D$ before it reaches its maximum~\cite{brandes2005}.
The radiated energy current can be derived from the time-derivative of the expectation value for
the energy.
To observe these effects, we initialize the system in the state of highest Hamming-distance ($\f{z}^0=\{0, \cdots ,0,1\}$).
The original Dicke effect refers to zero temperature, i.e., we consider the limit $\beta\to\infty$ here.
In this limit, we can easily solve Eqs.~(\ref{Eclasshamming}) and~(\ref{Equanthamming}) analytically
by e.g. Laplace transform~\cite{arfken2005}.
The solution of a system of the form
\bea
\dot y_0 &=& \beta_0 y_0 + \gamma_0 y_1\,,\nn
&\vdots&\nn
\dot y_i &=& \beta_i y_i + \gamma_i y_{i+1}\,,\nn
&\vdots&\nn
\dot y_n &=& \beta_n y_n
\eea
subject to the initial condition $\f{y}^0=(0,\ldots,0,1)$ is then given by
\bea\label{Esolgeneral}
y_k(t) = \left\{\sum_{a=k}^n e^{+\beta_a t} \left[\prod_{b=k : b\neq a}^n \frac{1}{\beta_a-\beta_b}\right]\right\}
\left[\prod_{c=k}^{n-1} \gamma_c\right]\,,
\eea
where we have assumed nondegenerate coefficients $\beta_a \neq \beta_{b\neq a}$.
In case of $\beta_a=\beta_b$ one may either use higher-order residue formulae or simply
analytic continuation of the above result.


\subsubsection{Rate Equations}

In the energy eigenbasis, we can easily relate the expectation value of the energy with the variables 
introduced in Eq.~(\ref{Evarhamrate}) via
\bea
\expval{E^{\rm re}} &=& \sum_{c=0}^{N-1} E_c \rho_{cc} = \sum_{\alpha = 0}^n E_{\alpha} \sum_{c \in {\cal H}_{\alpha}} \rho_{cc}\\
&=& \sum_{\alpha = 0}^n E_{\alpha} z_{\alpha}^{\rm re}(t)
= \omega_0\sum_{\alpha=0}^n \left(\alpha-\frac{n}{2}\right) z_\alpha^{\rm re}(t)\,.\nonumber
\eea
From Eq.~(\ref{Eclasshamming}) we identify in Eq.~(\ref{Esolgeneral}) in the zero-temperature limit
\mbox{$\beta_\alpha = - \alpha \lambda^2\eta^2 g(\omega_0)$} and
\mbox{$\gamma_\alpha=+(\alpha+1)\lambda^2\eta^2 g(\omega_0)$}.


\subsubsection{Quantum Master Equations}

\begin{figure}[ht]
\includegraphics[clip=true, width=0.45\textwidth]{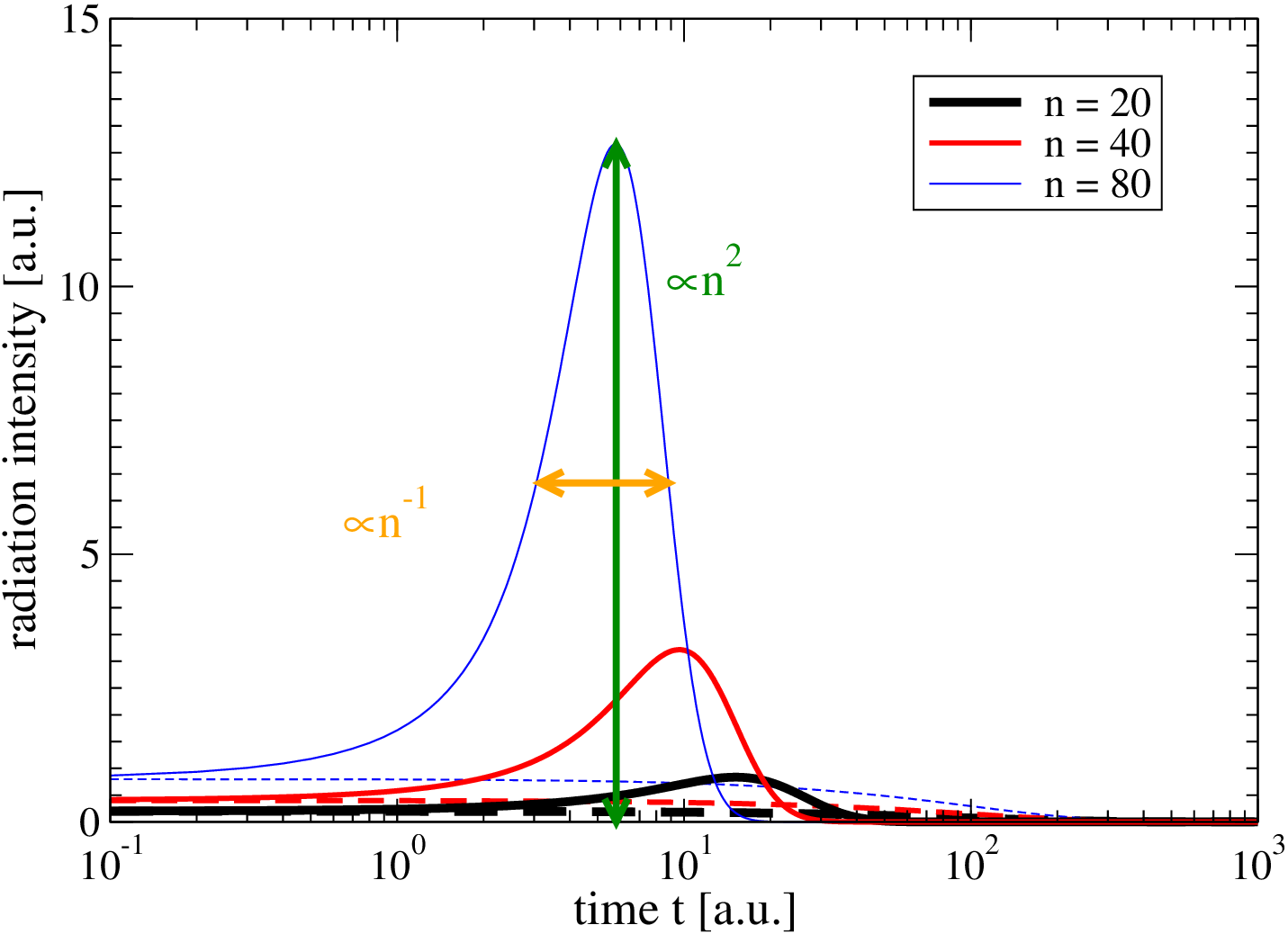}
\caption{\label{Fsuperrad}
[Color Online]
Plot of quantum master equation (solid) and rate equation (dashed) radiation intensity versus time for different system sizes $n$, 
with $n=20$ (black thick line), $n=40$ (red medium line) and $n=80$ (blue thin line).
Whereas in the incoherent case one always obtains simple exponential decay (dashed lines), the 
coherent case exhibits a strong flash with intensity proportional to $n^2$ and width proportional to $1/n$ after 
a delay time.
The totally radiated energy (area below the curves) is the same for both coherent and incoherent emission.
Other parameters have been chosen as $\beta \omega_0=100.0$, $\omega_0= 1.0$, $\lambda=0.1$, $g(\omega_0) = 1.0$.
}
\end{figure}

The quantum master equation expectation value of the energy can be obtained by noting that 
the Dicke Hamiltonian~(\ref{Ehamdicke}) is symmetric under all permutations of the spins, 
which implies that the whole evolution can be expressed with the angular momentum eigenstates 
$J^z \ket{j,m}=2 m \ket{j,m}$, where $j=n/2$.
These eigenstates relate to the superposition of all states in a given Hamming subspace via the relation
\bea
\ket{\frac{n}{2}, \alpha-\frac{n}{2}} = \frac{1}{\sqrt{\bin{n}{\alpha}}} \sum_{i \in {\cal H}_\alpha} \ket{i}\,,
\eea
where $\alpha \in \{0,1,2,\ldots,n\}$ denotes the Hamming distance to the ground state $\ket{1,\ldots,1}$.
For example, we see that for $n=4$ the states
\bea
\ket{2,-2} &=& \ket{1111}\,,\nn
\ket{2,-1} &=& \frac{1}{\sqrt{4}} \left[\ket{1110}+\ket{1101}+\ket{1011}+\ket{0111}\right]\,,\nn
\ket{2,\phantom{+}0} &=& \frac{1}{\sqrt{6}} \biggl[\ket{0011}+\ket{0101}+\ket{1001}\nn
&&+\ket{1100}+\ket{1010}+\ket{0110} \biggr]\,,\nn
\ket{2,+1} &=& \frac{1}{\sqrt{4}} \left[\ket{0001} + \ket{0010} +\ket{0100} +\ket{1000} \right]\,,\nn
\ket{2,+2} &=& \ket{0000}
\eea
are symmetric under all permutations of the qubits.
This can be exploited to relate the master equation expectation value for the energy with the
variables introduced in Eq.~(\ref{Evarhamquant}) via
\bea
\expval{E^{\rm qe}} &=& \frac{\omega_0}{2} \trace{J^z \rho}\nn
&=& \omega_0 \sum_{\alpha=0}^n \left(\alpha-\frac{n}{2}\right)\bra{\frac{n}{2},\alpha-\frac{n}{2}}\rho \ket{\frac{n}{2},\alpha-\frac{n}{2}}\nn
&=& \omega_0 \sum_{\alpha=0}^n \left(\alpha-\frac{n}{2}\right) \frac{1}{\bin{n}{\alpha}} \sum_{a,b\in {\cal H}_\alpha} \bra{a} \rho \ket{b}\nn
&=& \omega_0 \sum_{\alpha=0}^n \left(\alpha-\frac{n}{2}\right) \frac{z_\alpha^{\rm qe}(t)}{\bin{n}{\alpha}}\,.
\eea
From Eq.~(\ref{Equanthamming}) we identify in Eq.~(\ref{Esolgeneral}) for zero temperature the
coefficients
\mbox{$\beta_\alpha=- \alpha(n-\alpha+1) \lambda^2\eta^2 g(\omega_0)$} and
\mbox{$\gamma_\alpha=+ (\alpha+1)^2 \lambda^2\eta^2 g(\omega_0)$}.
These can be used to obtain the short-time dynamics from Eq.~(\ref{Esolgeneral}):
Obviously, we have $z_\alpha^{\rm qe} \le \bin{n}{\alpha}$, such that initially
only the variables for large $\alpha$ will contribute
\bea
\expval{E^{\rm qe}} &\approx& \omega_0 \left[ \frac{n}{2} z_n^{\rm qe} + \frac{\frac{n}{2}-1}{n}  z_{n-1}^{\rm qe}\right]\,.
\eea
Performing the time derivative and assuming large $n$,
we get an approximate expression for the radiation intensity
\bea\label{Erad_approx}
I &\approx& e^{-2 \lambda^2 g(\omega_0) n t} g(\omega_0) \lambda^2 n^2 \omega_0
\left(e^{+\lambda^2 g(\omega_0) n t} - 1\right)\,.
\eea
For this approximate expression, we can analytically extract the position and height of the maximum as well as the width 
of the radiation peak as
\bea
t_{\rm peak} &=& \frac{\ln(2)}{\lambda^2 g(\omega_0) n} \propto \frac{1}{n}\,,\nn
I_{\rm peak} &=& \frac{\omega_0}{4} \lambda^2 g(\omega_0) n^2 \propto n^2\,,\nn
\Delta_{\rm peak} &=& \frac{2 \ln \left[\sqrt{\frac{\sqrt{2}+1}{\sqrt{2}-1}}\right]}{\lambda^2 g(\omega_0) n} \propto \frac{1}{n}\,.
\eea
If one considers the uppermost three variables instead of the uppermost two, one obtains similar scaling relations.
These rough scaling relations for the Dicke flash are well reflected in the full solution, see Fig.~\ref{Fsuperrad},
where also the solutions for the rate equation radiation intensity are displayed for comparison.

\section{Conclusion}\label{Sconclusion}

We observe a clearly faster relaxation speed when coherences are considered, i.e., Markovian 
relaxation may be significantly slowed down by dephasing noise or -- if degeneracies are slightly lifted
-- in the later stages.
The heuristic explanation is, that coherences open more channels of transitions for the system 
to reach the groundstate. 
In the simple examples we have considered, this speed up goes along with a non-ergodic evolution of 
the quantum master equation, i.e., the stationary state depends on the initial state, whereas for
the rate equations evolution was always ergodic.
In order to restore ergodicity, more different coupling operators (decoherence channels) 
are generally needed~\cite{breuerpetruccione2002}.
Note that -- slight energy splittings provided -- ergodicity is automatically provided, since for 
times larger than the inverse level splittings the rate equation (with ergodic evolution) becomes
valid.
A smooth transition between quantum master and rate equations is given by the dynamical coarse-graining method~\cite{schaller2008a} 
or also found in the singular coupling limit~\cite{schultz2009a}.

For the non-local coupling operators considered, we did not obtain the algorithmic efficiency of the Grover search.
Note however, that this is possible in dissipative approaches, either phenomenologically 
(i.e., without deriving the Liouville superoperator from the interaction Hamiltonian)~\cite{spreeuw2007}
or also for an evolution under a Hamiltonian~\cite{mizel2009}.
In addition, not even the classical scaling was met by the rate equation approach for local operators.
However, this originates in our strict constraints on the coupling operator norm.

Furthermore, we note that the algorithms can be extremely efficient, when the matrix element of the
coupling operator between the initial state and the ground state $\bra{s}A\ket{w}$ is large.
When these matrix elements become independent of the system size $n$, we may also observe relaxation times 
that become independent of $n$.
These findings could be interesting for ground state quantum computation, where we can encode the
solutions to difficult problems in the ground state of a Hamiltonian.

When considering local operators, we have re-discovered the Dicke superradiance Hamiltonian as a 
special case, such that one may now generalize the superradiance not only to finite temperatures, 
but also towards non-equally distributed energy levels.
Note however, that in the latter case the permutational symmetry of the Dicke Hamiltonian is destroyed, 
such that the radiance will then have to be derived by means of an $n$-resolved master equation approach~\cite{cook1981a}.

\begin{appendix}

\section{Single Projector: Calculation of $\dot{\f{z}}$ for the quantum case}\label{Acoherenceone}

From Eq.~(\ref{Equantum}) we obtain
\bea
\dot \rho_{ww} &=& - \sum_{c \neq w} \bar\gamma_{cw,cw} \rho_{ww} + \sum_{cd \neq w} \bar\gamma_{wc,wd} \rho_{cd}\\
&=& - \frac{\lambda^2}{N^2} (N-1) \gamma(-\Delta E) \rho_{ww} + \frac{\lambda^2}{N^2} \gamma(+\Delta E) \sum_{ab \neq w}\rho_{ab}\,,\nonumber
\eea
which is coupled to $\sum_{ab \neq w}\rho_{ab}$. Thus we also need the time-evolution for $ z_1 = \sum_{ab \neq w} \rho_{ab}$ which can be calculated to
\bea
\dot{z_1} &=&  \sum_{ab \neq w} \sum_{c \neq w}\left[-i \bar\sigma_{ca} +i \bar\sigma_{bc}-\frac{1}{2}\bar\gamma_{wa,wc} -\frac{1}{2}\bar\gamma_{wc,wb}\right] \rho_{ab}\nn
&&+ \sum_{ab \neq w} \sum_{cd \neq w} \left[\bar\gamma_{ca,db} -\frac{1}{2}\bar\gamma_{ca,cd} -\frac{1}{2}\bar\gamma_{cd,cb}\right] \rho_{ab}\nn
&&+ \sum_{ab \neq w} \bar\gamma_{aw,bw} \rho_{ww} \nn
&=& - \frac{\lambda^2}{N^2} (N-1) \gamma(\Delta E) \sum_{ab \neq w} \rho_{ab} \nn
&&+ \frac{\lambda^2}{N^2} (N-1)^2 \gamma(-\Delta E) \rho_{ww}\,.
\eea
The Lamb shift terms cancel and the system of equations is closed, such that we
obtain Eq.~(\ref{eq:sym_quant}).


\section{Hadamard Coupling: Calculation of $\dot{\f{z}}$ for the quantum case}\label{ArelaxII}

Note that we make extensively use of
\bea
\frac{1}{N} &&\sum_c (-1)^{a \circ c + b \circ c} = \delta_{ab}\,,\nn
\frac{1}{N} &&\sum_{c\neq w} (-1)^{a \circ c + b \circ c} = \delta_{ab} - \frac{1}{N} (-1)^{a\circ w + b \circ w}\,,
\eea
which is derived from the property
\bea
A^2 = \bigotimes_{\ell=1}^n \frac{(\sigma^x_\ell + \sigma^z_\ell)^2}{2} = \f{1}\,.
\eea
Therefore, one obtains
\bea
   \sum_c \bra{a}A \ket{c}\bra{c}A\ket{b} &&= \sum\limits_c \frac{1}{N} (-1)^{a \circ c + c \circ b} \nn
					  &&= \bra{a} A \ket{b} = \delta_{ab}\,.
\eea
The time-evolution of $z_0=\rho_{ww}$ is with Eq.~(\ref{Equantum}) given by
\bea
\dot \rho_{ww} &=& - \left[\sum_{c\neq w} \bar \gamma_{cw,cw}\right] \rho_{ww} + \sum_{cd\neq w} \bar\gamma_{wc,wd} \rho_{cd}\nn
&=& -\lambda^2 \gamma(-\Delta E) \frac{N-1}{N} \rho_{ww} \nn
&&+ \lambda^2 \frac{\gamma(+\Delta E)}{N} \sum_{ab\neq w} (-1)^{w\circ a + w \circ b} \rho_{ab}\,,
\eea
whereas the time-evolution for $z_1(t)$ is given by the following, where $\sigma_{ij}$ and $\gamma_{ij,kl}$ were inserted:
\bea
\dot z_1 &=& -i \frac{\lambda^2}{2i} \frac{\sigma(+\Delta E)}{N} \sum_{abc\neq w} \bigl[ (-1)^{w\circ c + w \circ b} \rho_{cb}\nn
         && - (-1)^{w\circ a + w\circ c} \rho_{ac}\bigr]\nn
&&-i \frac{\lambda^2}{2i} \frac{\sigma(0)}{N} \sum_{abcd\neq w}\bigl[ (-1)^{d\circ a + d\circ c + w\circ a + w\circ b} \rho_{cb}\nn
&&- (-1)^{d\circ c + d\circ b + w\circ a + w\circ b} \rho_{ac}\bigr]\nn
&&+\lambda^2 \gamma(-\Delta E) \frac{(N-1)^2}{N} \rho_{ww} \nn
&&+ \lambda^2 \frac{\gamma(0)}{N} \sum_{abcd\neq w} (-1)^{w\circ a + w\circ b + a\circ c + b\circ d} \rho_{cd}\nn
&&- \frac{1}{2} \lambda^2 \frac{\gamma(+\Delta E)(N-1)}{N} \sum_{bc\neq w} (-1)^{w\circ c + w \circ b} \rho_{cb}\nn
&&- \frac{1}{2} \lambda^2 \frac{\gamma(0)}{N} \sum_{abcd\neq w} (-1)^{w\circ a + w \circ b+c\circ d + c\circ a} \rho_{db}\nn
&&- \frac{1}{2} \lambda^2 \frac{\gamma(+\Delta E)(N-1)}{N} \sum_{ac\neq w} (-1)^{w\circ a + w \circ c} \rho_{ac}\nn
&&- \frac{1}{2} \lambda^2 \frac{\gamma(0)}{N} \sum_{abcd\neq w} (-1)^{w\circ a + w \circ b+c\circ b+c\circ d} \rho_{ad}\,.
\eea
This can be written as
\bea
\dot{z}_1&=& - \frac{\lambda^2}{2} \sigma(0) \sum_{abc\neq w}\biggl[ \delta_{ac} (-1)^{w\circ c+w\circ b}- \frac{1}{N} (-1)^{w\circ a+w\circ b}\nn
&&-\delta_{bc} (-1)^{w\circ a + w\circ c} + \frac{1}{N} (-1)^{w\circ b+w\circ a}\biggr] \rho_{ab}\nn
&&+\lambda^2 \gamma(-\Delta E) \frac{(N-1)^2}{N} z_0 - \lambda^2 \frac{\gamma(+\Delta E)(N-1)}{N} z_1\nn
&&+ \lambda^2 \frac{\gamma(0)}{N} \left[1-\frac{1}{2} - \frac{1}{2}\right] z_1\,.
\eea
In the above equations, the Lamb-shift terms cancel and the system closes, such that
we obtain the results of Subsec.~\ref{quantum_eq}.


\section{Non-ergodicity: Overlap of solution with the steady state}\label{ap:overlap}
For an initial $\ket{s}\bra{s}$ density matrix, the initial variable $z_2$ in Eq.~(\ref{Evarhadamard}) equates to
\bea
z_2^0 &=& \frac{1}{N} \sum_{a,b\neq w} (-1)^{a \circ w + b \circ w}\nn
 &=& \frac{1}{N} \sum_{a \neq w} \prod_{l=1}^n(-1)^{a_l \circ w_l} \sum_{b \neq w} \prod_{k=1}^n(-1)^{b_k \circ w_k} \nn 
&=& \frac{1}{N} \left(\sum_{a_1 \in \left\{ 0,1\right\}} \cdots \sum_{a_n \in \left\{ 0,1\right\}} \prod_{l=1}^n (-1)^{a_l \circ w_l} - (-1)^{d_w} \right)^2 \nn
 &=& \frac{1}{N} \left(\prod_{l=1}^n \left\{(-1)^{0_l \circ w_l} + (-1)^{1_l \circ w_l} \right\} - (-1)^{d_w} \right)^2\nn
&=&\frac{1}{N} \left(\prod_{l=1}^n \left\{ 1 + (-1)^{w_l} \right\}-(-1)^{d_w} \right)^2\nn
&=& \begin{cases} \frac{1}{N} \left(N-1 \right)^2 \textmd{ for } \ket{w} = \ket{0 \cdots 0} \\
		  \frac{1}{N} \left(-1 \right)^{2 d_w} = \frac{1}{N} \textmd{ for } \ket{w} \neq \ket{ 0 \cdots 0} 
  \end{cases}\nn
&=& (N-2) \delta_{w,0\ldots 0} + \frac{1}{N},
\eea
where $d_w$ denotes the Hamming distance of the solution $\ket{\omega}$ to the state $\ket{0 \cdots 0}$.

\end{appendix}


\end{document}